\documentclass[twocolumn, showpacs, showkeys, amsmath, amssymb, prb, aps]{revtex4}
\usepackage{dcolumn}
\usepackage{bm}
\usepackage[dvips]{graphicx}
\begin{document}
\title{Characteristics of strong ferromagnetic Josephson junctions
with epitaxial barriers}
\author{C. Bell}
\altaffiliation{Present address: Kamerlingh Onnes Laboratory,
Universiteit Leiden, Leiden, The Netherlands. Email:
bell@physics.leidenuniv.nl} \affiliation{Materials Science
Department, IRC in Superconductivity and IRC in Nanotechnology,
University of Cambridge, Cambridge, UK}

\author{R. Loloee}
\affiliation{Department of Physics and Astronomy, Center for Sensor
Materials and Center for Fundamental Material Research, Michigan State
University, East Lansing, Michigan 48824, USA}

\author{G. Burnell, M. G. Blamire}
\affiliation{Materials Science Department, IRC in
Superconductivity and IRC in Nanotechnology, University of
Cambridge, Cambridge UK}
\date{\today}
\begin{abstract}
We present the measurement of superconductor / ferromagnetic
Josephson junctions, based on an epitaxial Nb bottom electrode and
epitaxial Fe$_{20}$Ni$_{80}$ barrier. Uniform junctions have been
fabricated with a barrier thicknesses in the range $2-12$ nm.  The maximum
critical current density $\sim 2.4 \pm 0.2 \times 10^9$ Am$^{-2}$
was found for a devices with a 3 nm thick barrier at 4.2 K,
corresponding to an average characteristic voltage $I_C R_N$ $\sim
16$ $\mu$V. The $I_C R_N$ showed a non-monotonic behavior with
Fe$_{20}$Ni$_{80}$ thickness. The variation of the
resistance of a unit area $AR_N$, of the junctions with barrier thickness gave a
Nb/Py specific interface resistance of $6.0
\pm 0.5$ f$\Omega$m$^2$ and Fe$_{20}$Ni$_{80}$ resistivity of $174 \pm 50$
n$\Omega$m, consistent with other studies in polycrystalline samples.
\end{abstract}
 \maketitle
\section{Introduction}
The early research into the proximity effect between
superconductors (S) and ferromagnets (F) concentrated on
measurements of the critical temperature $T_C$ and critical field
of S/F heterostructures. A motivating factor for this research was
the realization of the $\pi$ state, in which the groundstate phase
difference between S layers was changes from 0 to $\pi$, due to
the oscillation of the superconducting order parameter induced in
the F layer.\cite{buzdin1982} The transition should manifest itself as a
non-monotonic change in the properties of the multilayers, as a
function of F layer thickness $d_F$. Many epitaxial and
polycrystalline systems involving different materials were
investigated, using various growth techniques. Although
oscillatory $T_C (d_F )$ were observed in several experiments (see
reference \cite{izyREVIEW} for a review), these studies were
complicated by interface effects and `dead' magnetic layers at the
S/F interface which made the interpretation of the oscillatory
$T_C$ more difficult.\cite{aarts}

It was not until the measurement of current perpendicular to plane
(CPP) Josephson junctions with ferromagnetic barriers (S/F/S) that
the $\pi$-shift could be conclusively demonstrated. Such
$\pi$-junctions have been characterised as a function of temperature
and $d_F$,\cite{ryazanov1, kontos} using alloys whose composition
could be tuned to achieve an appropriately low Curie temperature
($T_M$), such that the $0-\pi$ crossover could be observed in a
easily realizable window of experimental phase space.  These
$\pi$-junctions have since been incorporated into various loop
geometries to further demonstrate the $\pi$ shift.
\cite{ryazanov2, bauer}

S/F/S junctions have also been fabricated with the relatively high
$T_M$ ferromagnets Ni,\cite{blum} Co,\cite{surgers} and composite
Co/Cu/Fe$_{20}$Ni$_{80}$ structures.\cite{bellapl} In these cases
the junctions are much more sensitive to the barrier properties
and $d_F$, and hence the $0-\pi$ crossover has not been
demonstrated in junctions with strong ferromagnetic barriers.

In all of the above CPP junctions, the S and F layers have been
polycrystalline and in the dirty limit. The realization of
epitaxial junctions in the clean limit may remove some of the
difficulties of measuring low $T_M$ alloy systems, (which are
sensitive to stoichiometry and harder to characterize
magnetically), as well as the sensitivity to $d_F$ of the high
$T_M$ barriers. Clean limit junctions are also expected to show a
novel non-sinusoidal current-phase relationship \cite{golubovRMP},
in contrast to recent measurements of S/F/S junctions with
Cu$_x$Ni$_{1-x}$ alloy barriers.\cite{frolov}

S/F/S $\pi$-junctions have been proposed as potential logic
elements in quantum computing circuits \cite{ioffe, mooij};
however the present critical current - resistance product ($I_C
R_N$) values are relatively small at present. The clean limit may
again provide a route to achieving much higher critical current
densities $J_C$, and hence $I_C R_N$. The combination of all of
these factors motivates the investigation of S/F/S Josephson
junctions based on epitaxial ferromagnetic barriers. The epitaxial
Fe$_{20}$Ni$_{80}$ (Py) system is also of interest in spintronic
applications, such as the fabrication of ballistic spin valves and
spin torque devices, and also provides an interesting
comparison to the previous studies of polycrystalline  Nb/normal
metal (N) and F/N interfaces (for example \cite{yang, park,
holody}).

\section{Film growth and characterization}
Our $(1\overline{1}\,0)$ Nb/$(111)$ Fe$_{20}$Ni$_{80}$ (Py) films
are grown by sputtering on $(11\overline{2}\,0)$ Al$_2$O$_3$ as
described in detail elsewhere.\cite{loloee} To improve the epitaxy
and reduce the strain in the films, the actual device structure is
Nb/Cu/Py/Cu/Nb with the thicknesses of the two Cu layers $\sim$ 5
nm, (with $(111)$ orientation). The Cu layers are expected to be
strongly proximitized by the Nb electrodes and should not
significantly reduce the $J_C$ of the devices. The Py thickness
$d_{\mathrm{Py}}$, was in the range $2-12$ nm. The samples with
$d_{\mathrm{Py}} = 2, 4$ and 6 nm were initially grown with a
bottom Nb layer of thickness 200 nm, and a top electrode of 20 nm
of Nb followed by a 5 nm Au capping layer. This Au was removed
{\it ex-situ} by Ar ion milling, and the top Nb electrode
deposited by further d.c. sputtering. All of the remaining samples
were grown in a second deposition, with the top and bottom Nb
thicknesses of $\sim 250$ nm, deposited {\it in-situ}. To achieve
epitaxy, the Py barrier was grown at 423 K, and the Nb at 1023 K.
It was not possible therefore to grow the top Nb electrode
epitaxially at the reduced temperatures required for the Cu and Py
layers.
\begin{figure}[h]
\includegraphics[width=8.5cm]{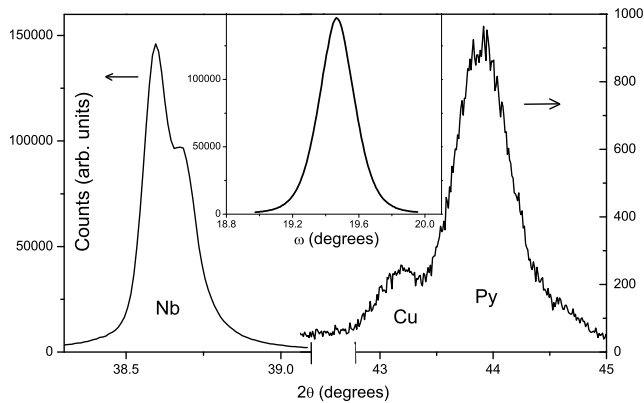}
\caption{\label{XRD}X-ray diffraction scan of $(1\overline{1}\,0)$
Nb, (111) Cu and (111) Py peaks at positions $2\theta = 38.6$,
43.2 and $43.9^{\circ}$ respectively. The splitting of the Nb peak
is caused by the presence of the two CuK$_{\alpha}$ radiation lines. Inset:
An $\omega$ scan of the primary Nb peak.}
\end{figure}
Fig. \ref{XRD} shows the x-ray diffraction peaks of the Nb, Cu and
Py layers in the $d_{\mathrm{Py}} = 6$ nm sample, (taken with
CuK$_{\alpha}$ radiation using a Philips X'Pert powder
diffractometer). The full width half maximum values obtained from
$\omega$ scans were $\sim 0.25^{\circ}$ for the Nb layer (inset of
Fig. \ref{XRD}), and $\sim 0.78^{\circ}$ for the Py layer (not
shown). This confirms the epitaxial nature of the bottom electrode
and barrier. A resistance {\it vs} temperature, $R(T)$,
measurement of the unpatterned $d_{\mathrm{Py}} = 2$ nm film
(before further Nb was deposited, such that the relatively thick
epitaxial Nb layer dominates the conductivity) was also made. The
film showed a residual resistance ratio $R(T = 10 \mathrm{K}) /
R(T = 300 \mathrm{K}) = 14.2$. This is of similar order to other
epitaxial Nb films, in which the superconducting coherence length
$\xi_S = 18$ nm, \cite{muhge1998} compared to a typical dirty
limit value of $\sim 6$ nm in polycrystalline sputtered
films.\cite{muhge, bellprb}

\begin{figure}[h]
\includegraphics[width=8.5cm]{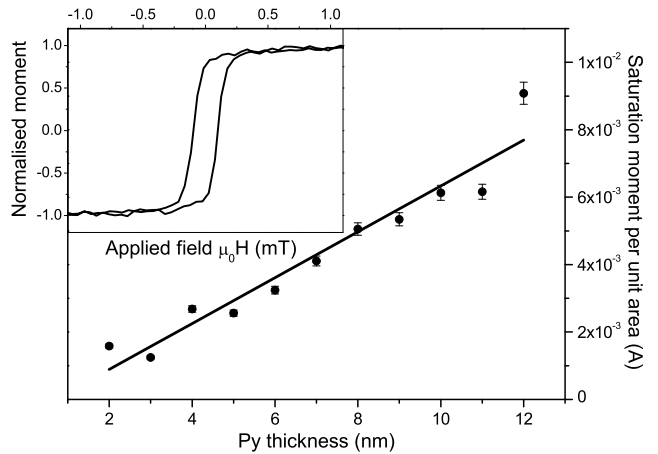}
\caption{\label{RTvsm}Scaling of the magnetic moment per unit area
vs Py thickness. The line is a least squares fit to the data.
Inset: Hysteresis loop of the 2 nm Py film at 295 K.}
\end{figure}
The magnetic properties were characterized with a vibrating sample
magnetometer at room temperature. Fig. \ref{RTvsm} shows the
saturation moment per unit area of the films in this study, (along with a
typical hysteresis loop shown in the inset). Extrapolating the
least square fit gives a nominal magnetically `dead' layer of
thickness $\sim 7$ \AA.

The films were patterned using optical lithography, followed by
broad beam Ar ion milling (1 mAcm$^{-2}$, 500 V) to micron scale
wires and associated tracks and contact pads, to allow four point
measurements to be performed on the devices. These tracks were
then processed in a Ga$^+$ focused ion beam to achieve vertical
transport with a device area in the range $0.05-1.1$ $\mu$m$^{2}$.
This fabrication process is described in detail elsewhere
\cite{bellnano} and has been used previously to fabricate
Josephson junctions with strong ferromagnetic
barriers.\cite{bellapl} Transport measurements were made in a
liquid He dip probe. The differential resistance as a function of
bias current of the junction was made with a lock-in amplifier,
and the $I_C$ found using a resistive criterion. The $R_N$ was
measured using a quasi-d.c$.$ bias current of $3-5$ mA. This
enabled the non-linear part of the $I-V$ near to $I_C$ to be
neglected, but was not too large to drive the Nb electrodes normal.

\section{Results and discussion}
\begin{figure}[h]
\includegraphics[width=8.5cm]{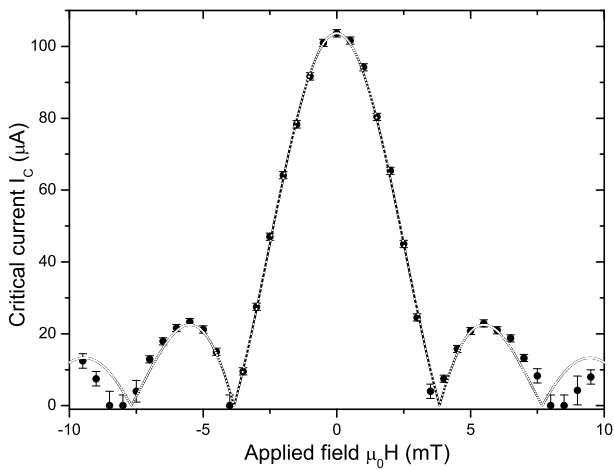}
\caption{\label{2nmich}$I_C (H)$ modulation for a device
demagnetized at 4.2 K for a 2 nm thick Py barrier. Device
dimension in the direction perpendicular to the applied field
$\sim$ 1.06 $\mu$m. Line is a best fit to a `Fraunhofer'
function.}
\end{figure}

An $I_C (H)$ modulation obtained in a junction with lateral area
$\sim$ 1060 nm $\times$ 250 nm is shown in Fig. \ref{2nmich}. In
this case the field is applied in the direction perpendicular to
the larger dimension of the device. It is expected that the
coercive field of the Py should increase relative to that taken
from the room temperature hysteresis loop shown in Fig.
\ref{RTvsm}, due to the reduced temperature and the aspect
ratio of the sub-micron patterned device.\cite{adeyeye} In this
case the applied field required to modulate the $I_C$ does not
significantly affect the magnetization of the Py, and the $I_C
(H)$ is symmetric about zero field. The good fit to the ideal
`Fraunhofer' pattern indicates a uniform current flow in the
junction. Relatively smaller junctions, which require larger
fields to modulate the $I_C$, were found to show
hysteresis which we associate with changes in the Py domain
structure and magnetization. Net magnetic induction present in the
barrier is known to shift the
$I_C (H)$ pattern away from $H = 0$ and reduce the $I_C$ at zero
field from its true maximum value. \cite{ryazanovquant}
To avoid falsely small values of $I_C$, the films were
demagnetized at 4.2 K, as well as the $I_C (H)$ pattern being
directly measured where possible, (for devices with $I_C R_N
> 2$ $\mu$V). In some cases however, the offset of the maximum $I_C$ from $H=0$ could
not be removed. The
cause of this is ascribed to shape anisotropy in the sub-micron
devices which provides an additional demagnetizing field which may
prevent the formation of a flux-closed domain structure in the
barrier.
\begin{figure}[h]
\includegraphics[width=8.5cm]{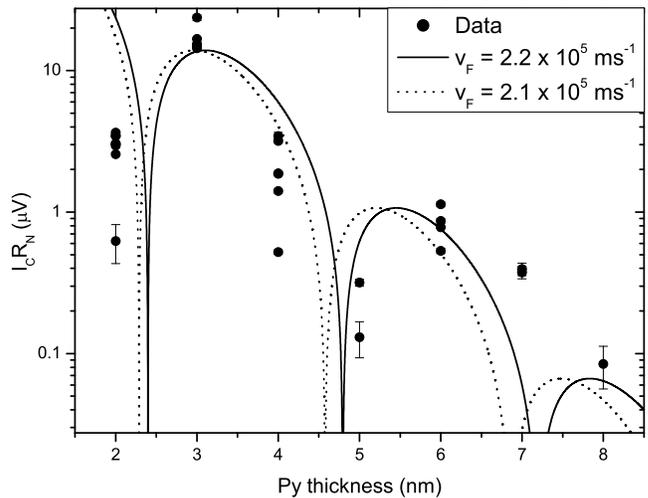}
\caption{\label{icrn}Characteristic voltage $I_C R_N$, as a
function of Py thickness at $T = 4.2$ K. Dashed and solid lines
are two fits to Eq. (\ref{sincfit}), as described in the text.}
\end{figure}

The variation of $I_C R_N (d_{\mathrm{Py}})$ is shown in Fig.
\ref{icrn}. Other devices showed much larger $J_C$ values, but
showed strongly distorted or no $I_C (H)$ modulation, implying
shorting around the edges of the junctions due to redeposited
material during device fabrication. For $d_{\mathrm{Py}} > 8$ nm
the devices showed some reduction of the differential resistance
around zero current bias, but did not show a measurable
supercurrent at $T = 4.2$ K, (this was also the case in several
devices with $d_{\mathrm{Py}} = 7$ and 8 nm). No re-entrant $I_C R_N$
was observed up to $d_{\mathrm{Py}} = 12$ nm. It is clear that
there is a strong suppression of $I_C R_N$ for increasing
$d_{\mathrm{Py}}$, but despite some scatter in the data of Fig. \ref{icrn} for each
set of devices with constant $d_{\mathrm{Py}}$, the decay is
clearly not purely exponential. A
component of this non-monotonic change may be associated with the
slightly different preparation methods of the 2, 4 and 6 nm thick
barriers, or run to run variation in the system. For example for
$d_{\mathrm{Py}}= 3$ nm, the fully {\it in-situ} deposition may
imply a higher quality and larger $I_C R_N$, however this is
inconsistent with the same comparison between the samples with
$d_{\mathrm{Py}} = 4, 5$ and 6 nm. Therefore the variation in $I_C
R_N$ would seem to be a true effect associated with the
oscillatory induced superconducting order parameter in the Py
layer.

We can compare the behavior of these junctions to the
non-epitaxial evaporated junctions of Blum {\it et al.}
\cite{blum}, with the structure Nb/Cu/Ni/Cu/Nb. In
that case a measurable critical current at $T = 4.2$ K was
observed up to a Ni thickness of 9 nm - similar to the present
data. Due to the relatively small number of data points and the
scatter, it is difficult to accurately fit this non-monotonic
decay. We model the data using
\begin{equation}\label{sincfit} I_C R_N \propto |\sin (2E_{\mathrm{ex}} d_F / \hbar
v_F ) | / (2E_{\mathrm{ex}} d_F / \hbar v_F )\; , \end{equation}
where $v_F$ is the Fermi velocity of Py, and $E_{\mathrm{ex}}$ the
exchange energy.\cite{blum, buzdin1982} These two parameters have
recently been measured\cite{petrovykh} in Py as $E_{\mathrm{ex}} = 135$ meV and
$v_F = 2.2 \pm 0.2 \times 10^5$ ms$^{-1}$ (for the majority spin).
The lines in Fig. \ref{icrn} correspond to $E_{\mathrm{ex}} = 95$
meV, with both $v_F = 2.2$ and $2.1 \times 10^5$ ms$^{-1}$, to
indicate the degree of variation the error in $v_F$ causes. The
agreement between the data and the model is not ideal, however for
an order of magnitude estimate it is clear that the fit is
acceptable. The period of oscillation of $I_C R_N (d_F)$ is given
by $\pi \hbar v_F / E_{\mathrm{ex}}$, and can therefore be
estimated to be of the order of 5 nm, again similar in magnitude
to the 5.4 nm value obtained by Blum {\it et al.}\cite{blum}

For the previously mentioned Ni junctions\cite{blum}, a maximum
$I_C \sim 20$ mA was observed at $T= 4.2$ K in $10 \times 10$
$\mu$m$^2$ devices for a 1 nm thick Ni barrier, giving $J_C = 2
\times 10^8$ Am$^{-2}$. In our samples, one device with
$d_{\mathrm{Py}} = 3$ nm had $I_C = 1.03$ mA with a lateral area
of $0.35 \pm 0.02$ $\mu$m$^2$, giving the highest $J_C \sim 2.9
\pm 0.2 \times 10^9$ Am$^{-2}$ at the same temperature,
the average for $d_{\mathrm{Py}} = 3$ nm was $J_C = 2.4 \pm
0.2 \times 10^9$ Am$^{-2}$. The larger $J_C$ for the epitaxial films can
be attributed to the increased
$\xi_S$ of the Nb bottom electrode.

\begin{figure}[h]
\includegraphics[width=8.5cm]{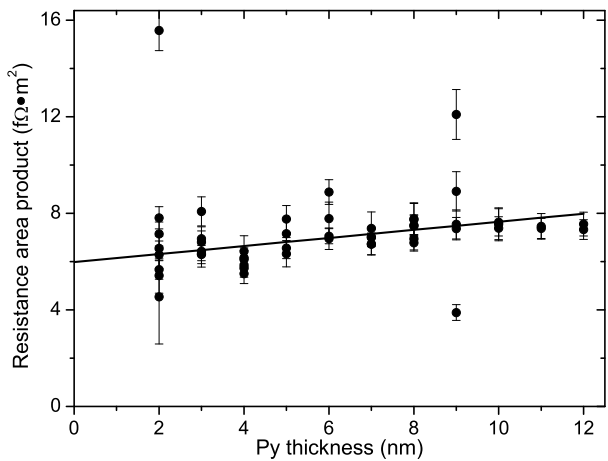}
\caption{\label{RAprod}Resistance area product ($A R_N$) at $T = 4.2$
K as a function of Py thickness. Line is a best fit excluding the point with
the largest $A R_N$ for $d_{\mathrm{Py}}= 2$ nm.}
\end{figure}

Finally, we have measured the total resistance ($R_N$) of a unit
area $AR_N$, of all junctions.  The variation of $AR_N$ for all
devices vs. $d_{\mathrm{Py}}$ over the range of 2
nm $< d_{\mathrm{Py}} <$ 12 nm are shown in Fig. \ref{RAprod}.
Here $AR_N$ is the total specific resistances; consisting of the
specific resistance of the S/F interfaces ($AR_{\mathrm{Nb/Py}}$)
in the S/F/S sandwich and the ferromagnetic layer,\cite{yang} such
that
\begin{equation} AR_N = 2AR_{\mathrm{Nb/Py}} +
\rho_{\mathrm{Py}} d_{\mathrm{Py}}.\end{equation}
If we exclude
one of the data points (with the highest $AR_N$ for
$d_{\mathrm{Py}} = 2$ nm), the best fit straight line to our data
gives an ordinate intercept of $2AR_{\mathrm{Nb/Py}} = 6.0 \pm
0.5$ f$\Omega$m$^2$ and the resistivity $\rho_{\mathrm{Py}} = 174
\pm 50$ n$\Omega$m (from the slope). The values fall within the
range of $2AR_{\mathrm{Nb/Py}} = 6-7.5$ f$\Omega$m$^2$ and $\rho_{\mathrm{Py}} =
110-140$ n$\Omega$m reported for polycrystalline samples
elsewhere.\cite{yang, holody}

\section{Summary}
We have measured the Josephson current through a epitaxial Py
barrier and observed high quality junction characteristics.
The $I_C R_N (d_{\mathrm{Py}})$ showed a non-monotonic
behavior, which could be approximately modelled with a simple
model. We observed no sign of re-entrant behavior above
$d_{\mathrm{Py}} = 8$ nm. The data extracted from the $AR_N(d_{\mathrm{Py}})$ product was
consistent with polycrystalline samples.

While interesting for spin-valve junctions, Py is not an
ideal material in which to fully explore the properties of
ferromagnetic Josephson junctions in the clean limit since the
spin diffusion length is relatively short.\cite{park, petrovykh} A
reduced spin diffusion length can have a strong influence
on the possible realization of $\pi$-junctions formed by the
interference of multiple Andreev reflection processes at the S/F
interfaces. The r{\^o}le of different reflection amplitudes for
the minority and majority spins, and the spin polarisation have been considered
elsewhere.\cite{barash,cayssol} Other effects on the Josephson current due
to shape of the Fermi surface in the epitaxial barrier may also be
a consideration, in an analogous fashion to the requirements in
tunnel junctions.\cite{dowman} The use of elemental, or other
epitaxial F layers - with longer spin diffusion lengths - may
allow even further increases of the $J_C$ to be achieved in these
structures, and rule out any problems associated with a loss of
spin memory. With a suitable choice of material and growth
technique, the fully epitaxial S/F/S would also be an important
system to study.

\section{Acknowledgements}
We thank J. Aarts for pointing out the importance of the spin
diffusion length in these systems, as well as E. J. Tarte, C. W.
Leung, W.P. Pratt and J. Bass for useful discussions and
assistance. We acknowledge the support of the Engineering and
Physical Sciences Research Council UK, the U.S. NSF grant
$98-09688$ and the European Science Foundation $\pi$-shift
network.

\end{document}